\documentstyle[amsfonts,aps,twocolumn]{revtex}

\newcommand{\p}{\partial}

\def\[{\left\lbrack}
\def\]{\right\rbrack}

\def\({\left(}
\def\){\right)}

\begin{document}

\title{The $SU(2)$ Skyrme model and anomaly\thanks{This work was partially supported 
by FAPEMIG, a Brazilian Research Agency.}}
\author{E. M. C. Abreu $^a$\thanks{Financially supported by Funda\c{c}\~ao de Amparo \`a 
Pesquisa do Estado de S\~ao Paulo (FAPESP).}, 
J. Ananias Neto and W. Oliveira${}^{b}$ }
\address{${}^{a}$ Departamento de F\'\i sica e Qu\'\i mica, Universidade Estadual Paulista, \\
Av. Ariberto Pereira da Cunha 333, Guaratinguet\'a, 12500-000, S\~ao Paulo, SP, Brazil, \\
e-mail: everton@feg.unesp.br \\
${}^{b}$ Departamento de F\'{\i}sica, ICE,  Universidade Federal 
de Juiz de Fora, \\ Juiz de Fora, $36036-330$, Minas Gerais, MG, Brazil, \\
e-mail: jorge@fisica.ufjf.br and wilson@fisica.ufjf.br}

\date{\today}

\draft

\maketitle

\begin{abstract}
\noindent The $SU(2)$ Skyrme model,expanding in the collective coordinates variables, 
gives rise to second-class constraints. Recently this system was embedded in a more general 
Abelian gauge theory using the BFFT Hamiltonian method.  In this work we quantize this gauge 
theory computing the Noether current anomaly using for this two different methods: an 
operatorial Dirac first class formalism and the non-local BV quantization coupled with the 
Fujikawa regularization procedure. 
\end{abstract}
\pacs{PACS: 11.10.Ef; 11.15.-q; 12.39.Dc; 11.30.Ly}

\section{Introduction}

The field-antifield formalism, created by I. Batalin and G. Vilkovisky (BV method) \cite{BV} 
has 
been used successfully to quantize the most difficult gauge theories such as supergravity theories and 
topological field theories in the Lagrangian framework \cite{GPS,Jon,Hen}. 
The BV method comprises the 
Faddeev-Popov quantization \cite{FP} and has the BRST symmetry as fundamental principle 
\cite{BRST}.  The method has 
introduced the definition of the antifield which are the sources of the BRST transformation, 
i.e., for each field we have an antifield canonically conjugated in terms of the antibracket 
operation.  With the fields, the antifields and the BRST transformation we can construct the 
classical BV action.  A mathematical ingredient, called the antibracket, helps 
us to construct the fundamental equation of the formalism at the classical level, the so-called 
master equation.  

At the quantum level we can define another mathematical operator, the $\Delta$ operator, which 
is a second order differential operator.  With the classical BV action and presenting local 
counterterms, we can construct the quantum BV action and analogously to the classical case, 
the quantum master equation.

The quantization is performed via the usual functional integration through the definition of the generating functional with the help of the well known Legendre transformation with the respect to the sources $J_A$. 
When it is not possible to find a solution to the quantum master equation we can say that the theory has an anomaly.  The presence of a $\delta(0)$ term in the $\Delta$ operation demand a method to treat this divergence conveniently.  
There are various methods to regularize the theory such as Pauli-Villars\cite{Pauli}, BPHZ \cite{BPHZ,Jonghe} and 
dimensional regularization \cite{T}.  
Recently, the non-local regularization \cite{Kle,Woo} 
coupled with the field-antifield formalism \cite{Paris,PT,eu} 
has been developed.  The success of the last one is based on its power to compute the anomaly of higher-loops.   The BRST superspace formulation brings another construction of the main 
ingredients of BV formalism \cite{ABD}.

The Skyrme model was first proposed by T. H. R. Skyrme \cite{Skyrme} in the sixties to 
incorporate baryons in the non-linear sigma model description 
of the low-energy interactions of pions \cite{P}.  A quantum theory can be constructed through the 
definition of the physical states which are annihilated by operators of the first-class 
constraints, and then we can obtain the physical values after taking the mean value of the canonical operators.  

Some efforts have been performed to quantize the Skyrme model.  Two of us \cite{ON} have achieved this by 
applying the non-abelian BFFT \cite{ON,BFFT} formalism, and thus, employ the Dirac method of 
first-class constraints to quantize the system \cite{Dirac}.  The analysis of the physical 
spectrum as well as the study of a hidden symmetry over the ordering ambiguity problem in 
the Skyrme model is discussed in \cite{CC}.  A formulation of the model as an embedded gauge
theory with the constraint deformed away from the spherical geometry is proposed in \cite{NW}.

Generally the anomaly is related to the fermionic aspect 
of the Skyrme-soliton physics or with 
the model coupled to fermions.  The first-class bosonic model has the auxiliary fields firstly introduced by L. D. Faddeev \cite{LDF} to convert a second-class system in a first-class one.  In \cite{GP10}, the idea of adding extra degrees of freedom has been implemented in the BV scheme.  Furthermore in \cite{JST}, it was shown that the cohomology of the classical theory has not been changed by the introduction of these new degrees of freedon.  As a consequence, the anomaly has not disappeared, but is shifted to these extra symmetries.   Hence we can say that the anomaly is in fact hidden in this way.

The purpose of this paper is to compute, in a precise way, 
the value of the 
anomaly of the bosonic model using an operatorial Dirac first-class formalism and
 the non-local field-antifield formalism coupled with the 
Fujikawa procedure of regularization \cite{Fuji}. Finally we want to introduce some ideas
about the geometry involved in the Skyrme model's anomaly.
The paper 
is organized as follow: in section 2,  the first class Skyrme 
action was derived and the operatorial calculation of the anomaly is developed; 
we have made a brief review of the non-local BV method in section 3;  
in section 4, we compute the one-loop anomaly.  The final 
considerations have been made in section 5.

\section{The Skyrme model and the operatorial method}

The classical static Lagrangian of the Skyrme model\cite{Skyrme} is given by

\begin{eqnarray}
\label{clag}
L &=& \int d^3r \{ -{F_\pi^2\over 16} Tr \(\partial_i U
\partial_i U^+ \) \nonumber \\
&+& {1\over 32 e^2} 
Tr \[ U^+\partial_i U, U^+ \partial_j U \]^2 \} \, ,
\end{eqnarray}

\noindent where $F_\pi$ is the pion decay constant, {\it e}
is a dimensionless parameter and U is an SU(2) matrix.
Performing the collective semi-classical expansion \cite{ANW},
substituting U(r) by $U(r,t)=A(t)U(r)A^+ (t)$ in (\ref{clag}),
where A is an SU(2) matrix, we obtain

\begin{equation}
\label{Lag}
L = - M + \lambda Tr [ \partial_0 A\partial_0 A^{-1} ],
\end{equation}

\noindent where M is the soliton mass, which in the hedgehog
representation for U, $U=\exp(i\tau \cdot \hat{r} F(r))$, 
is given by
\begin{eqnarray}
\label{henergia}
M &=& 4\pi {F_\pi \over e} \int^\infty_0 dx \, x^2 \left\{ {1\over 8}
\( F'^2 + {2 \sin^2 F \over x^2} \)\,+\,  \right. \nonumber \\
&+& \left. {1\over 2}
\[ {\sin^2 F\over x^2} \({\sin^2 F \over x^2}\,+\, 2 F'^2\) \]\, \right\} ,
\end{eqnarray}

\noindent where {\it x} is a dimensionless variable defined
by $x=eF_\pi r$ and $\lambda$ is called the inertia moment
written as $\lambda = {2 \over 3} \pi (e^3 F_\pi)^{-1} \Lambda$ with

\begin{equation}
\label{Lambda}
\Lambda = \int^\infty_0 dx\, x^2 \sin^2F \[ 1 +
4(F'^2 + {\sin^2 F\over x^2}) \].
\end{equation}

\noindent The SU(2) matrix A can be written as $A=a^0
+i a\cdot \tau$ with the constraint

\begin{equation}
\label{pri}
T_1 = a^ia^i - 1 \approx 0, \,\,\,\, i=0,1,2,3.
\end{equation}

\noindent The Lagrangian(\ref{Lag}) can be written as a function of the $a^i$ as

\begin{equation}
\label{lcol}
L = -M + 2\lambda \dot{a}^i\dot{a}^i.
\end{equation}

Batalin, Fradkin, Fradikina and Tyutin \cite{BFFT} developed an elegant formalism 
of transforming systems with second class constraints in first class ones, i.e., 
in gauge theories. This is achieved with the aid of auxiliary fields that extend
the phase space in a convenient way to transform the second class into first class 
constraints. This formalism and its extension non-Abelian \cite{Banerjee} were 
used for transforming the $SU(2)$ Skyrme model in an Abelian and non-Abelian gauge 
theory \cite{ON}, respectively. Since a Abelian gauge theory, the corresponding Lagrangian is 
given by

\begin{equation}
\label{equation}
\label{exl}
\tilde{L} = -M + 2\lambda {\dot{a}^i\dot{a}^i\over a^ia^i}
-2\lambda {\dot{\phi}\dot{\phi}\over (a^ia^i)^2},
\end{equation}
where $\phi^i$ are the auxiliary fields of BFFT formalism.

The new set of first class constraints are given by
\begin{eqnarray}
\label{nsc}
\tilde{T}_1 &=& T_1 + 2\phi, \\
\tilde{T}_2 &=& T_2 - a^ia^i \pi_\phi.
\end{eqnarray}

\noindent  The action is written as
 
\begin{equation}
\label{lag}
S = \int dt \left[-M+ 2\lambda {\dot{a}^i\dot{a}^i\over a^ia^i }
-2\lambda {\dot{\phi}\dot{\phi}\over {(a^ia^i)}^2} \right]\,\,,
\end{equation}

\noindent which is invariant for the following gauge transformations

\begin{eqnarray}
\label{cons1}
\delta a^j = \tau a^j,\\
\delta \phi = 2\tau \phi,
\end{eqnarray}

\noindent where $\tau$ is a constant parameter or a function dependent on position. The Noether current is obtained by using the formula\cite{HT}

\begin{equation}
\label{formula1}
j_0 = {\delta L\over \delta \dot{a}^i} \delta a^i + 
{\delta L\over \delta \dot{\phi}} \delta \phi \,\,,
\end{equation}

\noindent which result in

\begin{equation}
\label{noether2}
j_0=\  a^i\pi^i + 2\phi\pi_\phi\,\,.
\end{equation}

\noindent In the first class Dirac quantization constraints method \cite{Dirac}, the physical wave functions are obtained by imposing the condition

\begin{equation}
\label{qope}
\tilde{T}_\alpha | \psi \rangle_{phys} = 0, \,\,\,\, \alpha=1,2,
\end{equation}

\noindent being the operators $\tilde{T}^ 1 $ and 
$\tilde{T}^2 $ given by

\begin{eqnarray}
\label{qope1}
\tilde{T}^1=a^ia^i-1+2\phi,\\
\label{qope2}
\tilde{T}^2=a^i\pi^i - a^i a^i\pi_\phi.
\end{eqnarray}

\noindent Then, the physical states that satisfy (\ref{qope}) are
\begin{eqnarray}
\label{physical}
& & | \psi \rangle_{phys} = \nonumber \\
&=&  \, \delta (a^i\pi^i
-a^ia^i\pi\phi ) \,\delta(a^i a^i-1+2\phi)\,|polynomial \rangle,
\end{eqnarray}

\noindent where the ket {\it polynomial} is defined as
$|polynomial \rangle ={1\over N(l)} (a^1+ i a^2)^l \,$,
and $N(l)$ is a normalization factor. The polynomial kets
are the typical skyrmion collective coordinates 
eigenstates\cite{ANW}. 

Taking the scalar product, 
$_{phys} \langle \psi| j_0 | \psi \rangle_{phys}$, 
that is the mean value of the Noether current, formula (\ref{noether2}), we have
\begin{eqnarray}
\label{jmedio}
& &_{phys} \langle \psi| j_0 | \psi  \rangle_{phys} = \nonumber\\
&=& \langle polynomial |\,\, \int d\phi \,d\pi_\phi \cdot \nonumber \\
& & \cdot \, \delta(a^i a^i - 1 + 2\phi)\delta(a^i\pi^i - a^ia^i\pi_\phi)\, \cdot \nonumber \\ 
& & \cdot \, j_0  \delta(a^i a^i - 1 + 2\phi)\delta(a^i\pi^i - a^ia^i\pi_\phi)| polynomial \rangle . 
\end{eqnarray}

\noindent Note that due to $\delta(a^i a^i - 1 + 2\phi)$ and
$\delta(a^i\pi^i - a^ia^i\pi_\phi)$ in (\ref{jmedio}), the scalar product can be simplified. 
Then, integrating over $\phi$ and $\pi^\phi$ we obtain
\begin{eqnarray}
\label{jmedio2}
_{phys} \langle \psi| &j_0& | \psi \rangle_{phys} = \nonumber\\
&=& \langle polynomial | {a^i\pi^i\over a^ia^i} |polynomial \rangle . 
\end{eqnarray}

\noindent The operator $\pi^j$ describes a free momentum particle and its representation on the collective coordinates space $a_i$ is given by

\begin{equation}
\label{momentum}
\pi_j = -i {\partial\over\partial a_j}.
\end{equation}

\noindent The current operator inside the ket (\ref{jmedio2})
must be symmetrized. 
For this we use the Weyl ordering prescription \cite{Weyl}. This rule expresses that the new current operator must be constructed by counting all possible randomly order of the $a^i$,
$\pi^i$ and $1 / {(a^ia^i)}$. Then, we can write the symmetric current as
\begin{equation}
\label{jmw}
\[ {a^i\pi^i\over a^ia^i} \]_{sym} = {1\over a^ia^i}a^j\pi^j 
- {i\over a^i a^i},
\end{equation}  
where the ordering of the operator $a^j\pi^j/(a^ia^i)$ means that $a^j\pi^j$ acts firstly in the physical kets. The polynomial
states are eigenstates of the operator $a^i\pi^i$, i.e.,
$a^i\pi^i |polynomial \rangle =l |polynomial \rangle $. Thus,
the mean values of $j_0$ is given by\footnote{The regularization
of delta function squared like $\delta
(a^i a^i - 1 + 2\phi)^2$ and 
$\delta(a^i\pi^i - a^ia^i\pi_\phi)^2$ 
is performed by using the delta relation, 
$(2\pi)^2\delta(0)=\lim_{k\rightarrow 0}\int d^2x 
\,e^{ik\cdot x} =\int d^2x= V.$ }
\begin{eqnarray}
\label{jmf}
& &  _{phys} \langle \psi| j_0 | \psi \rangle_{phys} = \nonumber \\ 
&=&{-i(l+1)\over V^2}\langle polynomial | {1\over a^ia^i}|polynomial \rangle 
\nonumber \\
&=&{-i(l+1)\over V^2} {1\over a^ia^i}.
\end{eqnarray} 
\noindent In the above expressions, $a^ia^i$ is the square of the three-sphere collective 
coordinates radius. Then, the polynomial eigenstates do not depend of 
the term $1/{(a^ia^i)}$.  Moreover, the term $1 / {(a^ia^i)}$
in the definition of the mean value of $j_0$, eq. (\ref{jmf}), is an ``curvature scalar '' 
of the hypersurface defined by the second class constraint $T_1$, eq. (\ref{pri}) and, at first,
 in the first class system, it is not conserved in time.  This fact can indicate a possible 
anomaly in the Noether current $j_0$.

The current anomaly can be calculated by an scalar product given by
\begin{eqnarray*}
\label{dj0}
{\cal A} &=& \int \,dt\,_{phys} \langle \psi| \partial_0 j_0 | \psi \rangle_{phys} \\
&=&{1 \over V^2} \int dt \langle polynomial |\partial_0 \[ {a^i\pi^i \over a^ia^i} 
\]_{sym}|polynomial \rangle \\
&=& {-i(l+1) \over V^2} \int dt \langle polynomial | \partial_0 
{(1/ a^ia^i)} |polynomial \rangle.
\end{eqnarray*}

\noindent Using the definition of the momentum obtained
through the Lagrangian written in the action(\ref{lag}),
$\dot{a}^i = {a^ja^j \over 4\lambda} \pi^i $,
the anomaly ${\cal A}$ can be written as
\begin{eqnarray*}
\label{dj02}
{\cal A} = \frac{i\,(l\,+\,1\,)}{2\,\lambda\,V^2}
\int dt \langle polynomial |\[{a^i\pi^i\over a^ia^i}\]_{sym}
|polynomial \rangle. 
\end{eqnarray*}

\noindent Symmetrized the operator $ {a^i\pi^i/(a^ia^i)}$
written in the expression above, using for
this again the Weyl ordering prescription, and considering
that the ket {\it polynomial} is an eigenstate of the operator
$a^i\p^i$, we can finally obtain the expression for the anomaly ${\cal A}$, given by
\begin{equation}
\label{final}
{\cal A} =  \frac{(l\,+\,1\,)^2}{2\,\lambda\,V^2}\,
\int\,dt {1\over a^ia^i}\;\;,
\end{equation}
and from this expression we can see that the constant terms  have
to be computed.  Next we will do this using the BV nonlocal method finding the whole expression of this anomaly.

\section{The Field-Antifield Formalism with Non-local Regularization}

Let us construct the complete set of fields, ${\Phi^{A}}$,  
including in this set the classical fields, the ghosts 
for all gauge
symmetries and the auxiliary fields.  We can also define the antifields ${\Phi_{A}^{*}}$,
which are the canonical conjugated variables with respect to the antibracket structure,

\begin{equation}
(X,Y)= \frac{\delta_{r} X}{\delta \phi}\,\frac{\delta_{l}
Y}{\delta\phi^{*}}\, - \,( X \longleftrightarrow Y )\;\;,
\end{equation}

\noindent where the indices $r$ and $l$ denote, as usual, right and left functional derivatives respectively.

One can then construct an extended action of ghost
number equal to zero, the so-called BV action, also called classical proper solution,
 \begin{eqnarray}
\label{proaction}
S(\Phi,\Phi^{*}) &=& 
S_{cl}(\Phi) + \Phi^{*}_{A} R^{A}(\Phi) +
\frac{1}{2}\Phi^{*}_{A}
\Phi^{*}_{B} R^{BA}(\Phi)  \nonumber  \\
 &+& \ldots +\, \;\frac{1}{n!}
\Phi^{*}_{A_{1}} \ldots \Phi^{*}_{A_{n}} R^{A_{n}\ldots
A_{1}} +\ldots\;\;,
\end{eqnarray}

\noindent where $R^{A}(\Phi)$ are the BRST generators.  This action has to satisfy the classical master
equation, $(S,S)=0$. At the quantum level the action can be defined  by $W = S + \sum^{\infty}_{p=1} \hbar^{p}\,M_{p}$,
where the $M_{p}$ are the corrections (the Wess-Zumino terms)
to the quantum action. The quantization of the theory is obtained with the generating functional of the Green 
functions:

\begin{equation}
Z(J,\Phi^{*}) = \int
{\cal D} \Phi \,exp \, \frac{i}{\hbar} \left[ W (\Phi,
\Phi^{*}) + J^{A}
\Phi^{*}_A \right] .
\end{equation}

\noindent But the definition of a path integral properly 
lacks on a regularization framework.  

For a theory to be free of anomalies, the quantum action $W$ has to be a solution of the QME,
$\label{qme}(W,W) = 2\,i\,\hbar\, \Delta\,
W$, where $\Delta \equiv (-1)^{A+1}
\frac{\partial_{r}}{\partial\Phi^{A}}
\frac{\partial_{r}}{\partial\Phi^{*}_{A}}$.
In the QME one can see that when it is not possible to
find a solution, we have an anomaly that can be defined
by

\begin{equation}
\label{11}
{\cal A} \,\equiv \,\left[ \Delta W +
\frac{i}{2\hbar} (W,W) \right] (\Phi, \Phi^{*})\;\;.
\end{equation}

To accomplish the regularization we will choose a method developed recently \cite{Kle,Woo} 
and that has been adapted to the BV formalism \cite{Paris,PT}
\footnote{\noindent For convenience, in this quite brief review, we are using the same notation as the
reference \cite{Paris}.}. Let us define an action $S(\Phi)$ 
as being, $S(\Phi)=F(\Phi)+I(\Phi)$,
where $F(\Phi)$ is the kinetic part and $I(\Phi)$ is
the interacting part, 
which is an analytic function in $\Phi^{A}$ around
$\Phi^{A} = 0$. 
Then one can write conveniently that 
$F(\Phi)\,=\, \frac{1}{2} \Phi^{A} {\cal F}_{AB} \Phi^{B}$,
where ${\cal F}_{AB}$ is called the kinetic
operator. To perform the NLR we have
now to introduce a cut-off or regulating parameter
$\Lambda^{2}$.  An arbitrary and invertible matrix $T_{AB}$ has to be introduced too.  The
combination of ${\cal F}_{AB}$ with 
$(T^{-1})^{AB}$ defines a second
order derivative regulator, ${\cal R}^{A}_{B} = (T^{-1})^{AC} {\cal F}_{CB}$.

We can construct two important operators with these
objects.  The first is the smearing operator
$\epsilon^{A}_{B} = exp
\left( \frac{{\cal R}^{A}_{\;\;B}}{2\Lambda^{2}} \right)$,
and the second is the shadow kinetic operator

\begin{equation}
{\cal O}_{AB}^{-1} =
T_{AC}(\tilde{{\cal O}}^{-1})^{C}_{\;\;B} = 
\left( \frac{{\cal
F}}{\epsilon^{2} - 1} \right)_{AB}\;\;,
\end{equation}

\noindent with $(\tilde{{\cal O}})^{A}_{B}$ defined as

\begin{equation}
\tilde{\cal O}^{A}_{\;\;B}  =  \left(
\frac{\epsilon^{2} - 1}{{\cal R}} \right)^{A}_{\;\;B} 
 =  \int_{0}^{1} \frac{dt}{\Lambda^{2}}\, 
exp \left( \,t\,
\frac{{\cal R}^{A}_{B}}{\Lambda^{2}} \right)\;\;.
\end{equation}

\noindent In order to expand our original configuration space for each field $\Phi^{A}$, an
auxiliary field $\Psi^{A}$ can be constructed,  

\begin{equation}
\label{auxaction}
\tilde{{\cal S}}(\Phi,\Psi) \,=\,
F\,(\hat{\Phi}) \,-\, A\,(\Psi) \,+\, I\,(\Phi + \Psi)\;\;.
\end{equation}

\noindent The second term of this auxiliary action is called the auxiliary kinetic term,
$A(\Psi) \,=\, \frac{1}{2}\Psi^{A} ({\cal O}^{-1})_{AB} \Psi^{B}$.
The fields $\hat{\Phi}^{A}$, the smeared fields,
which make part of the auxiliary action are defined
by $\hat{\Phi}^{A} \equiv (\epsilon^{-1})^{A}_{\;\;B}
\Phi^{B}$.

In the non-local BV formalism the configuration space has to be enlarged
introducing the antifields $\{\Psi^{A},\Psi_{A}^{*}\}$.
Note that the
shadow fields have antifields too.  Then, an auxiliary
proper solution, $\tilde{S}(\Phi,\Phi^{*};\Psi,\Psi^{*})$,
incorporates the auxiliary
action (\ref{auxaction}) (corresponding to the
gauge-fixed action
$S(\Phi)$), its gauge symmetry and the
unknown
associated higher order structure functions.  The auxiliary
BRST
transformations are modified by the presence of the
term
$\Phi^{*}_{A}\,R^{A}(\Phi)$ in the original proper solution.
The shadow fields have to be substituted by the solutions of
their classical equations of motion.  At the same time, their
antifields will be equal to zero.  In this way we can write
$S_{\Lambda}(\Phi,\Phi^{*})=\tilde{S}(\Phi,\Phi^{*};\Psi,\Psi^{*} = 0)$,
and the classical equations of motion are
${\delta_{r}\,\tilde{S}(\Phi,\Phi^{*};\Psi,\Psi^{*})}/
{\delta
\Psi^{A}} = 0$
with solutions $\bar{\Psi} \equiv
\bar{\Psi}(\Phi,\Phi^{*})$.

The complete interaction term, ${\cal
I}(\Phi,\Phi^{*})$, of the original
proper solution can be written
as

\begin{equation}
{\cal I}(\Phi,\Phi^{*}) \equiv I(\Phi) \,+\,
\Phi^{*}_{A}\,R^{A}(\Phi) \,+\,
\Phi^{*}_{A}\,\Phi^{*}_{B}\,R^{BA}(\Phi)
\,+ \dots
\end{equation}

\noindent and the quantum action $W$ can be expressed by 
$W=F+{\cal I}+\sum^\infty_{p=1} \hbar M_p
\equiv F+{\cal Y}$,
where ${\cal Y}$ is the generalized quantum interaction part.
It can be shown that the
expression of the anomaly is the value of
the finite part in the limit
$\Lambda^{2} \longrightarrow \infty$ of 

\begin{equation}
\label{anomalia}
{\cal A}
= \left[ (\,\Delta\,W\,)_{R} \,+\,
\frac{i}{2\,\hbar}\,(W,W)
\right]\,(\Phi,\Phi^{*})
\end{equation}

\noindent and the regularized value of $\Delta W$ is defined as $(\Delta W)_{R} \equiv \lim_{\Lambda^{2}
\rightarrow \infty} \left[\Omega_{0}\right]$
where $\Omega_{0}$ and 
$\left( \delta_{\Lambda} \right)^{A}_{\;\;B}$ are defined by 
\begin{eqnarray}
\Omega_{0} &=& \left[
S_{\;\;B}^{A}\,\left(
\delta_{\Lambda} \right)^{B}_{\;\;C}\, \left( \epsilon^{2} \right)_{\;\;A}^{C}
\right], \nonumber \\
(\delta_{\Lambda})_{\;\;B}^{A}  &=&  \left( \delta^{A}_{\;\;B} -
{\cal
O}^{AC}\,{\cal I}_{CB} \right)^{-1} \nonumber \\
&=& 
\delta^{A}_{\;\;B} + \sum_{n=1}\, \left( {\cal O}^{AC}\,{\cal
I}_{CB}
\right)^{n}\;\;,
\end{eqnarray}

\noindent with $S^{A}_{\;\;B} =
{\delta_{r}\,\delta_{l}\,S}/{\delta\,\Phi^{B}
\,\delta\,\Phi^{*}_{A}}\,,\;
{\cal I}_{AB} = 
{\delta_{r}\,\delta_{l}\,{\cal
I}}/{\delta\,\Phi^{A}\,\delta\,\Phi^{B}}\,$. Applying
the limit $\Lambda^{2} \longrightarrow \infty$ in 
$(\Delta W)_{R}$, it
can be shown that 
$\left( \Delta S\right)_{R} \equiv
\lim_{\Lambda^{2} \rightarrow \infty}
\left[ \Omega_{0}
\right]_{0}\;\;,$
\noindent and finally that
${\cal A}_{0} 
\equiv  \left( \Delta\,S \right)_{R} = 
\lim_{\Lambda^{2} \rightarrow \infty} \left[ \Omega_{0}
\right]_{0}$.
All the higher orders terms of the anomaly can be
obtained from
equation (\ref{anomalia}), but this will not be analyzed in this paper. It can be seen in\cite{PT}.

\section{The non-local field-antifield quantization of the Skyrme model}

The first-class action for the massless Skyrme model is, using (\ref{equation}),

\begin{equation}
\label{action7}
{\cal S}\,=\,\int\,dt\,\left[\frac{2\,\lambda\,\dot{a}^i\,\dot{a}^i}{a^i\,a^i}
\,-\,\frac{2\,
\lambda\,\dot{\phi}^i\,\dot{\phi}^i}{(\,a^i\,a^i\,)^2}
\right]\;\;,
\end{equation}

\noindent but we already know that the first-class tell us that $2 \phi=1-a^i a^i$ so that 
$\dot{\phi}=-a^i \dot{a}^i$.  Substituting this in (\ref{action7}) we have now that

\begin{equation}
\label{action2}
{\cal S}\,=\,\int\,
dt\,\left[\frac{2\,\lambda\,\dot{a}^i\,
\dot{a}^i}{a^i\,a^i}
\,-\,\frac{2\,
\lambda\,({a^i} \dot{a}^i)^2} {(\,a^i\,a^i\,)^2}\right]\;\;,
\end{equation}

\noindent This action, as we can easily see, has a problem of non-locality, which can be solved 
expanding the terms,


\begin{eqnarray}
\label{action3}
{\cal S}\,&=&\,\int\,
dt\,\left\{\frac{2\,\lambda\,\dot{a}^i\,
\dot{a}^i}{[1-(1-a^i\,a^i)]}
\,-\,\frac{2\,
\lambda\,({a^i} \dot{a}^i)^2} {[1-(1-\,a^i\,a^i)]^2}\right\} \nonumber \\
&=& 2 \lambda \dot{a}^i\,\dot{a}^i \sum_{n=0}^\infty (1-a^i\,a^i)^n\, \nonumber \\
&-& \,2 \lambda 
(\,a^i\,\dot{a}^i\,)^2 \sum_{n=0}^\infty (n+1)(1-a^i\,a^i)^n
\end{eqnarray}

\noindent This
action is invariant for the BRST transformations given by \cite{HKP}

\begin{equation}
\delta
a^i \, = \,c\,a^i\;,\:\:\:\:\:\:\:\:\:\:\:\:
\delta c\,= \, 0\;\;.
\end{equation}

\noindent Now we can construct the BV action,

\begin{eqnarray}
\label{actionBV}
{\cal S}_{BV}\,&=& \,  \int dt\, \left\{\, 2 \lambda \dot{a}^i\,\dot{a}^i \sum_{n=0}^\infty 
(1-a^i\,a^i)^n\, \right.\nonumber \\
&-& \left.\,2\, \lambda \,
(\,a^i\,\dot{a}^i\,)^2 \sum_{n=0}^\infty (n+1)(1-a^i\,a^i)^n \,+\,a^*_i c a^i\, \right\}
\end{eqnarray}

\noindent The kinetic part of the action (\ref{action7}) after an integration by parts 
is,

\begin{equation}
F  = \int d\,t\,[\,2\,\lambda\,\dot{a}^i\,\dot{a}^i\,] = \int dt [\,-2\,\lambda\,
a^i\,\p^2_0\, a^i\,]\;\;.
\end{equation}

\noindent The kinetic term has the form

\begin{equation}
F = \frac{1}{2} a^i (-4 \lambda \p^2_0 )\, a^i\;\; \Longrightarrow \;\; {\cal F}_{AB}\,=\,
-4 \,\lambda \,\p^2_0 \,\delta_{AB}
\end{equation}

\noindent The regulator, a second order differential operator, can be chosen as

\begin{equation}
{\cal R}^A_B = \p^2_0\;\;,\;\;\;\Longrightarrow\;\;T\,=\,-4 \lambda\;\;.
\end{equation}

\noindent where $T$ is an arbitrary non-singular matrix.
The smearing operator has the form,
$\epsilon^{A}_{\;\;\;B} = exp \left( \frac{\partial^{2}_0}{
2\,\Lambda^{2}} \right)
\,\delta^A_{\;\;\;B}$. In the NLR scheme the shadow kinetic operator is

\begin{equation}
{\cal O}_{AB}^{-1} =
\left( \frac{{\cal F}}{\epsilon^{2} - 1}
\right)_{AB}
= \left( \frac{-4 \lambda \p^2_0}{\epsilon^{2} - 1} \right)_{AB}
\end{equation}

\noindent where

\begin{equation}
{\cal O}^{AB} = -\,\frac{\epsilon^{2} - 1}{4\, \lambda \,\p^2_0}
\;=\;-\,\int^1_0 \frac{d \tau}{\Lambda^2}\, exp \left( \tau \frac{4\,\p^2_0}{\Lambda^2}\right)
\end{equation}

\noindent Using the definitions of $S^A_{\;\;B}$ and ${\cal I}_{AB}$ we can show that, 
\begin{eqnarray}
S^a_{\;\;a}\,&=&\,c \\
{\cal I}_{aa} &=& -\, 4 \,\lambda \,\p^2_0\;+\;\frac{4 \,\lambda \,\p^2_0}{a^i\,a^i}\;-\;
\lambda \frac{(16 a^i \dot{a}^i \p_0 \;+\;\dot{a}^i \dot{a}^i)}{(a^i\,a^i)^2} \nonumber \\
&-&\lambda \frac{( \dot{a}^i \dot{a}^i 
\,+\,12\,a^i \dot{a}^i\p_0\,+\,(a^i \p_0)^2)}{(a^i\,a^i)^2} 
\,+\,8 \lambda \frac{\dot{a}^i \dot{a}^i a^j\,a^j}{(a^i\,a^i)^3} \nonumber \\
&-& \, 8 \lambda \frac{(\dot{a}^i \dot{a}^i)^2}{(a^i\,a^i)^4}
\end{eqnarray}

\noindent Finally the anomaly can be computed as we know

\begin{eqnarray}
{\cal A}\,&=&\,(\Delta S)_R\, \nonumber \\
&=&\,\lim_{\Lambda^2 \rightarrow \infty} \{Tr[\epsilon^2 S^A_{\;\;B}]
\,+\,Tr[\epsilon^2 S^A_{\;\;D} {\cal O}^{DC} {\cal I}_{CB}] \}
\end{eqnarray}

\noindent Computing each term we have, for the first term 
in ${\cal I}_{aa}$, writing  only the main steps,


\begin{eqnarray}
& &
\lim_{\Lambda^2 \rightarrow \infty} 
\left[ -\,4\,\epsilon^2\,\lambda\,c\,\int\,dt \,{\cal O}
\,\p^2_0 \right] \nonumber\\
&=& 
\lim_{\Lambda^2 \rightarrow \infty} \[ -\,4\,\epsilon^2\,\lambda\,c\,\int\,dt \int 
\frac{dk}{2 \pi} e^{-ikt}\,{\cal O} \p^2_0 exp \left( \frac{\p^2_0}{\Lambda^2} \right) 
e^{ikt} \] \nonumber \\
&=& \lim_{\Lambda^2 \rightarrow \infty} \[ -\,4\,\epsilon^2\,\lambda\,c\,\int\,{dt \over 
\Lambda}\, \cdot \right.  \nonumber \\
&\cdot& \left. \,\int^1_0 \(- \frac{d \tau}{\Lambda^2} \)\, exp 
\left( \tau \frac{4\,\p^2_0}{\Lambda^2}\right) \cdot
\int \frac{dk}{2 \pi}(-\,k^2) exp \left( \frac{-k^2}{\Lambda^2} \right) \] \nonumber \\
&=& \lim_{\Lambda^2 \rightarrow \infty} \[ -\,4\,\epsilon^2\,\lambda\,c\,\int\,
{dt \over \Lambda}  \int d \tau \left( 1 + \frac{4\,\p^2_0}{\Lambda^2} \right)
\left( - {\sqrt{\pi} \over 2} \right) \] \nonumber \\
&=& \lim_{\Lambda^2 \rightarrow \infty} \[ -\,4\,\epsilon^2\,\lambda\,
\left( - {\sqrt{\pi} \over 2} \right)  \int\,dt \,c \]
\end{eqnarray}

\noindent where we have made two convenient
reparametrizations $(t,k) \rightarrow (\lambda t,\lambda k)$ to solve the integrals \cite{grad}.

Analogously for another term in ${\cal I}_{aa}$

\begin{eqnarray}
\lim_{\Lambda^2 \rightarrow \infty}& & \left\{ 3\,\epsilon^2\,\lambda\,c\,\int\,dt \,{\cal O}
\,\frac{\p^2_0}{a^i\,a^i} \right\}
\;=\; \nonumber \\
&=& \lim_{\Lambda^2 \rightarrow \infty} \left\{ {3 \over 2} \sqrt{\pi}\,\epsilon^2\,\lambda\,
\int\,dt  {c \over a^i\,a^i}  \right\} \;\;.
\end{eqnarray}

\noindent Doing the same procedure for all the other 
terms one can conclude that they are identically zero.

As we know, terms that depends only on ghots does not have any physical meaning in the final 
result of the anomaly.  Computing only the physical terms, 
the one-loop anomaly for the $SU(2)$ Skyrme model is the 
Wess-Zumino consistent expression \cite{WZ},

\begin{equation}
{\cal A} = {3 \over 2} \sqrt{\pi}\,\lambda\,
\int\,dt \, {c \over a^i\,a^i}\;\;.
\end{equation}

\noindent It is a new result, corroborating the general expression founded before in 
eq. (\ref{final}) showing an anomaly in the conservation of the Noether 
current $j_0$. But at the level of BV formalism, as well known, it represents an impossibility to the solution of the QME.

\section{Conclusions}

In this work we have considered the $SU(2)$ Skyrme model as a more general Abelian gauge 
theory. Firstly we have quantized this gauge theory using an operatorial Dirac first-class 
formalism and computing the anomaly of the Noether current as a manifestation of the Gaussian 
curvature of the hypersurfaces of constraints.  But this method yields an expression dependent 
on unknown geometrical constant terms.  It is the nonlocal field-antifield formalism which 
discloses the whole anomaly expression.  
 
\section{Acknowledgments}

The author EMCA would like to thank the hospitality
of the Departamento de F\'{\i}sica of the Universidade 
Federal de Juiz de Fora where part of this work was done and the financial support of
Funda\c{c}\~ao de Amparo \`a Pesquisa do Estado de S\~ao Paulo (FAPESP).

\end{document}